\documentclass[aps,pra,reprint,superscriptaddress,floatfix,hyperref]{revtex4-1}
\usepackage{amsmath,amssymb} 
\usepackage{graphicx}

\usepackage[breaklinks]{hyperref}

\begin{document}

\title{Non-linear mixing of Bogoliubov modes in a bosonic Josephson junction}

\author{Sof\'ia Mart\'inez-Garaot}
\affiliation{Departamento de Qu\'imica F\'isica, Universidad del Pais Vasco UPV/EHU, 48080 Bilbao, Spain}

\author{Giulio Pettini}
\affiliation{Dipartimento di Fisica e Astronomia, Universit\`a di Firenze,
and INFN, 50019 Sesto Fiorentino, Italy}

\author{Michele Modugno}
\affiliation{\mbox{Depto. de F\'isica Te\'orica e Hist. de la Ciencia, Universidad del Pais Vasco UPV/EHU, 48080 Bilbao, Spain}}
\affiliation{IKERBASQUE, Basque Foundation for Science, 48013 Bilbao, Spain}

\begin{abstract}
We revisit the dynamics of a Bose-Einstein condensate in a double-well potential, from the regime of Josephson plasma oscillations to the self-trapping regime, by means of the Bogoliubov quasiparticle projection method. 
For very small imbalance between left and right wells only the lowest Bogoliubov mode is significantly occupied. In this regime the system performs plasma oscillations at the corresponding frequency, and the evolution of the condensate is characterized by a periodic transfer of population between the ground and the first excited state. As the initial imbalance is increased, more excited modes -- though initially not macroscopically occupied  -- get coupled during the evolution of the system. Since their population also varies with time, the  frequency spectrum of the imbalance turns out to be still peaked around a single frequency, which is continuously shifted towards lower values. The nonlinear mixing between Bogoliubov modes eventually drives the system into the the self-trapping regime, when the population of the ground state can be transferred completely to the excited states at some time during the evolution. For simplicity, here we consider a one-dimensional setup, but the results are expected to hold also in higher dimensions.
\end{abstract}

\date{\today}

\maketitle

\section{Introduction}

Two weakly coupled Bose-Einstein condensates (BECs) in a double-well potential constitute a paradigmatic system for investigating the physics of bosonic Josephson junctions \cite{javanainen1986,milburn1997,smerzi1997,albiez2005}. Owing to the nonlinear character of the interactions, this system exhibits different dynamical behaviors, ranging from Josephson plasma oscillations (in the limit of very small imbalance between the population of the two wells) \cite{josephson1962}, to macroscopic self-trapping where -- above a critical value of the imbalance -- the population of the two wells is almost locked to the initial value \cite{smerzi1997,albiez2005,gati2006}. Due to the conceptual importance of these phenomena, BECs in double-well potentials and arrays of coupled boson Josephson junctions have been extensively investigated in the last two decades both theoretically \cite{dalfovo1996,milburn1997,smerzi1997,zapata1998,raghavan1999,giovanazzi2000,zhang2001,sakellari2002,ananikian2006,rosenkranz2008,giovanazzi2008,ichihara2008,julia-diaz2009,julia-diaz2010,ottaviani2010,julia-diaz2010a,julia-diaz2010b,mele-messeguer2011,wuster2012,julia-diaz2012,julia-diaz2012a,julia-diaz2013,jezek2013,li2013,burchianti2017} and experimentally \cite{cataliotti2001,albiez2005,anker2005,gati2006,gati2006b,gati2007,levy2007,leblanc2011,trenkwalder2016,labouvie2016,spagnolli2017}, as well as their counterparts with fermionic superfluid atomic samples \cite{spuntarelli2007,ancilotto2009,zou2014,valtolina2015}. 
 
 The physics of these systems is well captured by a two-mode approximation of the Gross-Pitaevskii equation, each mode being localized in one of the two wells, which allows for an effective description in terms of only two parameters, namely the population imbalance $z(t)$ and the phase difference $\phi(t)$ between the left and right components. Here we provide a complementary description by means of the quasiparticle projection method of Ref. \cite{morgan1998}, extending the Bogoliubov treatment of Ref. \cite{burchianti2017} to the case of arbitrary initial imbalance. For the sake of simplicity, we shall restrict the analysis to the case of a (quasi) one-dimensional condensate \footnote{For the definition of a (quasi) one-dimensional condensate see e.g. \cite{modugno2018} and references therein.}
 
We find that in the regime of a small initial imbalance, where only one Bogoliubov mode is significantly occupied and the system performs plasma oscillations at the corresponding frequency \cite{burchianti2017}, the evolution of the condensate is characterized by a periodic transfer of population between the ground state and the first excited state. As the initial imbalance is increased, more Bogoliubov modes get coupled during the evolution of the system, and their population also varies with time, contrarily to what happens in a linear system. As a consequence, the frequency spectrum of the imbalance turns out to be still peaked around a single frequency which is shifted towards lower values, rather than getting relevant contributions at higher frequencies, where Bogoliubov modes are located. By further increasing the initial imbalance, the population of the ground state can be completely transferred to the excited states at some time during the evolution, driving the system into the macroscopic self-trapping regime. 
 
The paper is organized as follow. In Sec. \ref{sec:model} we introduce the formalism, reviewing the definition of the two-mode approach (\ref{sec:2mode}) and of the quasiparticle Bogoliubov expansion (\ref{sec:bogol}). Then, in Sec. \ref{sec:results} we present the results by discussing the behavior of the system in the regime of Josephson plasma oscillations (\ref{sec:joseph}), the self-trapping regime (\ref{sec:st}), and that intermediate between the former two (\ref{sec:interm}), highlighting the role of non-linear mixing (\ref{sec:nl}).  Final considerations are drawn in the conclusions.
 
\section{Model}
\label{sec:model}

\begin{figure}[]
\includegraphics[width=0.8\columnwidth]{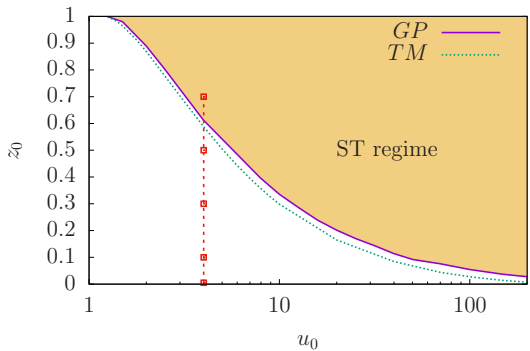}
\caption{Plot of the critical initial imbalance $z_{0c}$ vs $u_{0}$, as obtained from the solution of the GP equation (solid line). The shaded area indicates the self-trapping (ST) regime, for $\mu/V_{0}=0.25$. Empty squares on the vertical dashed line represent the values of $z_{0}$ and $u_{0}$ considered in this paper. The prediction of the TM model \cite{smerzi1997} in Eq. (\ref{eq:z0crit}) is also shown as a reference (dotted line).}
\label{fig:zcrit}
\end{figure}

Let us consider the following (dimensionless) Gross-Pitaevskii equation \cite{burchianti2017}
\begin{equation}
i\partial_{t}\psi(x,t)= \left[-\frac{1}{2}\nabla_{x}^{2}+V(x) + u_{0}|\psi(x,t)|^{2} \right]\psi(x,t)
\label{eq:gp}
\end{equation}
with
\begin{equation}
V(x)=\frac12{(x+\delta x)}^{2} + {V}_{0}{e}^{-2{x}^{2}/{w}^{2}},
\end{equation}
and $\int\!\! dx|\psi(x)|^2 = 1$, describing a (quasi) one dimensional condensate trapped in a double-well potential. The latter is composed by a harmonic potential term, plus a barrier of intensity $V_{0}$ and width $w$, with $\delta x$ providing a relative shift between the two. Here we are interested in describing the dynamics triggered by an initial population imbalance between the two wells. This can be obtained by preparing the system in the ground state $\psi_{g}(x)=\psi(x,0)$ of the above potential with $\delta x\neq0$, and then suddenly switching $\delta x=0$ at $t=0$ (only the harmonic potential is shifted, the barrier does not move). 
The ground state $\psi_{g}(x)$ is obtained from
\begin{equation}
\left[-\frac{1}{2}\nabla_{x}^{2}+V(x) + u_{0}|\psi_{g}(x)|^{2} \right]\psi_{g}(x)=\mu\psi_{g}(x),
\end{equation}
with $\mu$ being the condensate chemical potential.
As for the parameters, here we choose ${w}=0.3$ and ${V}_{0}=50$, that correspond to a double-well configuration within reach of current experiments (see e.g. Ref. \cite{valtolina2015}), whereas the interaction strength $ u_{0}$ and the initial shift $\delta x$ are taken as free parameters, and will be varied for exploring different regimes (see later on). In particular, $\delta x$, is chosen in order to produce the desired initial imbalance $z_{0}$. 

As it is known, in the limit of very small initial imbalance the system performs Josephson plasma oscillations \cite{javanainen1986}, and eventually enters a self-trapping (ST) regime at a critical imbalance \cite{smerzi1997} whose specific value depends on the strength $ u_{0}$ of the nonlinear term (see Fig. \ref{fig:zcrit}). 
The dynamics of the system will be analyzed by means of an expansion over the Bogoliubov modes, by comparing with the exact evolution and the two-mode (TM) approach.

\subsection{Two-mode model}
\label{sec:2mode}

Usually, the dynamics of a condensate in a double-well potential is treated by means of the two-mode approach, which consists in writing the condensate wave function as (see e.g. \cite{pitaevskii2016,burchianti2017} and references therein)
\begin{equation}
 \psi(x,t)=c_{L}(t)\psi_{L}(x)+c_{R}(t)\psi_{R}(x),
\end{equation}
where the functions $\psi_{L,R}(x)$ are localized in the left and right well, have unit norm, and are orthogonal to each other, $\langle\psi_{L}|\psi_{R}\rangle=0$. Though somewhat approximate, -- and not entirely justified from the formal point of view \cite{burchianti2017} -- the two-mode model provides an effective description of the double-well system in several respects, and will be used in the rest of the discussion as a reference. Here we construct the two-modes $\psi_{L,R}(x)$ from the ground state $\psi_{g}(x)$ (symmetric) and the first-excited solution $\psi_{1}(x)$ (antisymmetric) of the stationary GP equation 
\footnote{The stationary GP equation is solved by imaginary time evolution \cite{dalfovo1999}. Once the ground state $\psi_{g}$ has been obtained, the first-excited solution is found by the same approach, requiring it to be orthogonal to $\psi_{g}(x)$.}. Namely, we take the following linear combination, $\psi_{L/R}\equiv(\psi_{g}\pm \psi_{1})/\sqrt{2}$ \cite{pitaevskii2016}, 
corresponding to the most common approach in the literature \cite{raghavan1999,sakellari2004,danshita2005,ananikian2006,gati2007,adhikari2009,julia-diaz2010,pitaevskii2016}.
Then, by defining ($\alpha=L,R$)
\begin{equation}
 c_{\alpha}(t)=\sqrt{N_{\alpha}(t)}e^{\displaystyle{i\phi_{\alpha}(t)}},
\end{equation}
and 
\begin{eqnarray}
K&\equiv &-\int\!\! dx \psi_{\alpha}(x){\hat{H_{0}}}\psi_{\beta}(x)\nonumber\\
 U_{\alpha mn\beta} &\equiv& u_{0}\int\!\! dx \psi_{\alpha}(x)\psi_{m}(x)\psi_{n}(x)\psi_{\beta}(x)
\end{eqnarray}
one gets the following equations for the phase difference $\phi\equiv\phi_{\alpha}-\phi_{\beta}$ and the imbalance $z\equiv N_{\beta}- N_{\alpha}$ \cite{ananikian2006},
\begin{align}
\frac{\dot{z}}{2K}&=(2\Lambda_{1}-1)\sqrt{1-z^2}\sin \phi+(1-z^2)\Lambda_{2}\sin{2\phi},
\\
\frac{\dot{\phi}}{2K}&= (\Lambda-2\Lambda_{2})z
+\frac{(1-2\Lambda_{1})z}{\sqrt{1-z^2}}\!\cos\phi
-z\Lambda_{2}\cos2\phi,
\end{align}
with $\Lambda\equiv U_{\alpha\alpha\alpha\alpha}/2K$,
$\Lambda_{1}\equiv U_{\alpha\alpha\alpha\beta}/2K$, 
$\Lambda_{2}\equiv U_{\alpha\alpha\beta\beta}/2K$.
In the following, this set of equations will be referred to as the full two-mode (fTM) model. 
When the terms $\Lambda_{1}$ and $\Lambda_{2}$ can be neglected, it reduces to the well-known two-mode (TM) model by Smerzi \textit{et al.} \cite{smerzi1997}.

We recall that the TM model predicts that the system enters the ST regime when the parameter $\Lambda$ exceeds a critical value $\Lambda_{c}$. For $\phi_{0}=0$, it takes the following value:
$\Lambda_{c}(z_{0})=2(\sqrt{1-z_{0}^2}+1/z_{0}^2)$ \cite{smerzi1997}.
Here, we shall use as independent parameter $u_{0}$ rather than $\Lambda$ (which will depend on $u_{0}$). Then, the previous equation can be easily inverted, yielding
\begin{equation}
z_{0c}=\left(2\sqrt{\Lambda( u_{0})-1}\right)/\Lambda( u_{0}).
\label{eq:z0crit}
\end{equation}
As shown in Fig. \ref{fig:zcrit}, this formula provides a good estimate for the actual critical imbalance extracted from the GP equation, in the whole range considered ($u_0\in [1,200]$).

\subsection{Bogoliubov approach}
\label{sec:bogol}

As a complementary description, here we employ the quasiparticle projection method introduced by Morgan \textit{et al.} in Ref. \cite{morgan1998}. It amounts to a Bogoliubov expansion \cite{dalfovo1999,castin2001} where the condensate and quasiparticle populations are allowed to vary with time, namely
\begin{equation}
\psi(x,t)=e^{-i\mu t/\hbar}\left[\psi_{g}(x)(1+b_{g}(t))+\delta\psi(x,t)\right],
\label{eq:exp1}
\end{equation}
with
\begin{equation}
\delta\psi(x,t)=
\sum_{i>0}
b_{i}(t) \tilde{u}_{i}(x) + b^{*}_{i}(t) \tilde{v}^{*}_{i}(x).
\label{eq:exp2}
\end{equation}
The functions $\tilde{u}_{i}(x)$ and $\tilde{v}_{i}(x)$ are the Bogoliubov eigenmodes,
with the tilde indicating that they are chosen to be orthogonal to $\psi_{g}(x)$ \footnote{
In general, the solutions of Eq. (\ref{eq:bogol}) are not orthogonal to ground-state solution $\psi_{g}$. However, one has the freedom to impose that condition, posing \cite{morgan1998} 
${u}_{i}=\tilde{u}_{i}-a_{i}\psi_{g}$, $\tilde{v}_{i}^{*}=\tilde{v}_{i}^{*}+a_{i}^{*}\psi_{g}$,
with $a_{i}=\int \psi_{g}^{*}{u}_{i}=-\int \psi_{g}{v}_{i}$.}. They are
 solutions of (from now on we fix $\psi_{g}^{*}=\psi_{g}$ without loss of generality) \footnote{The Bogoliubov equations are transformed into matrix equations by grid discretization.}
\begin{equation}
\left(\begin{matrix}
 {\cal L}&g\psi_{g}^{2}\\
-g\psi_{g}^{2}&-{\cal L}\\
\end{matrix}
\right)\left(
\begin{matrix}
\tilde{u}_{i}\\
\tilde{v}_{i}\\
\end{matrix}
\right)= \omega_{i}
\left(
\begin{matrix}
\tilde{u}_{i}\\
\tilde{v}_{i}\\
\end{matrix}
\right),
\label{eq:bogol}
\end{equation}
with
\begin{equation}
{\cal L}\equiv -\frac{1}{2}\nabla_{x}^{2}+V(x) + 2g\psi_{g}^{2}-\mu.
\end{equation}
The solutions of Eq. (\ref{eq:bogol}) satisfy the following orthogonality relations \footnote{In general, there are three different classes of solutions of Eq. (\ref{eq:bogol}), with ``norm'' $\int \tilde{u}_{i}^{*}\tilde{u}_{j} - \tilde{v}_{i}^{*}\tilde{v}_{j}=\pm\delta_{ij},0$; only one of the two families with norm $\pm1$ has to be considered in the expansion (\ref{eq:exp2}) \cite{castin2001}.} 
\begin{align}
\int\!\!dx\left[\tilde{u}_{i}^{*}(x)\tilde{u}_{j}(x) - \tilde{v}_{i}^{*}(x)\tilde{v}_{j}(x)\right]&=\delta_{ij},
\\
\int\!\!dx\left[\tilde{u}_{i}(x)\tilde{v}_{j}(x) - \tilde{v}_{i}(x)\tilde{u}_{j}(x)\right]&=0.
\end{align}

The coefficients $b_{g}(t)$ and $b_{i}(t)$ are given by \cite{morgan1998}
\begin{align}
&b_{g}(t)=\int \!\!dx\!\left[\psi_{g}(x)\psi(x,t) e^{i\mu t}\right]-1,
\\
&b_{i}(t)=\int \!\!dx\!\left[\tilde{u}_{i}^{*}(x)\psi(x,t) e^{i\mu t}\!\! - 
\tilde{v}_{i}^{*}(x)\psi^{*}\!(x,t) e^{-i\mu t} 
\right].
\label{eq:boft}
\end{align}

In the \textit{linear regime}, when the modes remain decoupled during the whole evolution, the coefficients $b_{i}(t)$ are solutions of $i\dot{b}_{i}(t)=\omega_{i}b_{i}(t)$, namely \cite{morgan1998,dalfovo1999}
\begin{equation}
b_{i}(t)=b_{i0}e^{-i\omega_{i}t},
\label{eq:bilinear}
\end{equation}
where the coefficients $b_{i0}\equiv b_{i}(0)$ (which do not depend on time) are fixed by the initial conditions, see Eq. (\ref{eq:boft})
\begin{equation}
b_{i0}=\int\!\!dx\left[\tilde{u}_{i}^{*}(x) - \tilde{v}_{i}^{*}(x)\right]\psi_{g}(x).
\end{equation}

\textit{Imbalance.}
To construct the population imbalance between the right and left wells we start by integrating the particle density
\begin{align}
 &n(x,t)\equiv |\psi(x,t)|^{2}
\simeq
|1+b_{g}(t)|^{2}|\psi_{g}(x)|^{2}
\\
&\quad
+2Re\left[\psi_{g}(x)(1+b_{g}^{*}(t))
\sum_{i}\left(
b_{i}(t) \tilde{u}_{i}(x) + b^{*}_{i}(t) \tilde{v}^{*}_{i}(x)
\right)\right]
\nonumber
\end{align}
over the positive and negative $x$ semi-axis. By taking into account the symmetries of the problem we have
\begin{equation}
N_{R,L}(t)=A(t)\pm B(t)
\end{equation}
with 
\begin{align}
A(t)&=|1+b_{g}|^{2}\int_{0}^{+\infty}\!\!\!\!\!\!\!\!\!dx~|\psi_{g}|^{2}
\nonumber
\\
\nonumber
&+2Re\left[(1+b_{g}^{*})
\!\!\!\sum_{i\in even}\!\!\!\left(
b_{i}\!\! \int_{0}^{+\infty}\!\!\!\!\!\!\!\!\!dx~\psi_{g}\tilde{u}_{i} 
+ b^{*}_{i}\!\! \int_{0}^{+\infty}\!\!\!\!\!\!\!\!\!dx~\psi_{g}\tilde{v}^{*}_{i}
\right)\right],
\\
B(t)&\simeq 2Re\left[\left(1+b_{g}^{*}\right)\!\!
\sum_{i\in odd}\!\!\left(
b_{i}\!\! \int_{0}^{+\infty}\!\!\!\!\!\!\!\!\!dx~\psi_{g}\tilde{u}_{i} 
+ b^{*}_{i}\!\! \int_{0}^{+\infty}\!\!\!\!\!\!\!\!\!dx~\psi_{g}\tilde{v}^{*}_{i}
\right)\right].
\label{eq:bfull}
\end{align}
Then, the \textit{imbalance} $z(t)\equiv N_{R}(t)-N_{L}(t)$ is
\begin{equation}
z(t)=2B(t).
\label{eq:imbalance}
\end{equation}
Remarkably, only the Bogoliubov excitations with odd $i$ contribute to the imbalance, owing to the symmetries of the system.
In the \textit{linear regime} we have (using also the fact that in our case 
$\tilde{u}_{i}$, $\tilde{v}_{i}$ can be chosen real without loss of generality)
\begin{align}
B(t)&\simeq2\sum_{i\in odd}\left[b_{i0}\int_{0}^{+\infty}\!\!\!\!\!\!\!\!\!dx~\psi_{g}\left(\tilde{u}_{i} + \tilde{v}_{i}\right)
\right]\cos(\omega_{i}t)
\nonumber\\
&\equiv 2\sum_{i\in odd}B_{0i}\cos(\omega_{i}t).
\label{eq:linearcase}
\end{align}

\section{Results and discussion}
\label{sec:results}

\begin{figure}[]
\includegraphics[width=\columnwidth]{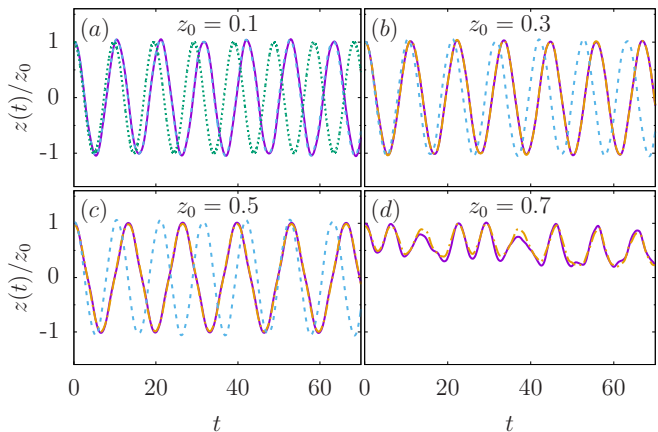}
\caption{Evolution of the normalized imbalance, $z(t)/z_{0}$, for $z_{0}=0.1$ (a), $0.3$ (b), $0.5$ (c), $0.7$ (d). The different lines correspond to the solution of the GP equation (solid purple line), the TM model [dotted green line, in (a)], and the prediction of the quasiparticle projection method in Eq. (\ref{eq:imbalance}), with $B(t)$ given by Eq. (\ref{eq:bfull}) [short-dashed orange line, in (b)-(d)] and Eq. (\ref{eq:linearcase}) [dot-dashed cyan line, in (a)-(c)].} 
\label{fig:fig2}
\end{figure}

Here we shall discuss the evolution of the imbalance for different values of its initial value $z_{0}\equiv z(0)$ (throughout this work we set $\phi_{0}\equiv\phi(0)=0$), discussing the behavior of the system in terms of the quasiparticle projection method \cite{morgan1998} introduced in the previous section. The general behavior of $z(t)$ has already been extensively studied, at least in the framework of the TM model, see e.g. the seminal Ref. \cite{smerzi1997}. 
In the rest of this paper we fix the ratio $\mu/V_{0}\equiv0.25$, a value that characterizes a typical Josephson regime (with the chemical potential much lower than the barrier height \cite{pitaevskii2016}).
The explicit behavior of the imbalance evolution is shown in Fig. \ref{fig:fig2}
for $u_0=4$ and $z_{0}=0.1,0.3,0.5,0.7$ (empty squares in Fig. \ref{fig:zcrit}),
ranging from the regime of Josephson plasma oscillations ($z_{0}\lesssim0.1$), to the ST regime ($z_{0}\gtrsim0.62$). A detailed description of the different dynamical behaviors and of the various lines plotted in the figure is given in the following.

\subsection{Josephson plasma oscillations} 
\label{sec:joseph}

In the limit of very small imbalance, the system performs Josephson plasma oscillations characterized by a frequency $\omega_{J}$. This frequency corresponds to the energy of the lowest Bogoliubov mode \cite{burchianti2017}. In fact, in this limit only one Bogoliubov mode is occupied, the system is in the \textit{linear regime}, and $z(t)$ is well reproduced by Eqs. (\ref{eq:imbalance}), (\ref{eq:linearcase}) with the only contribution of $B_{01}$, namely \cite{burchianti2017} 
\begin{equation}
z(t)=4B_{01}\cos(\omega_{1}t).
\label{eq:linearzeta}
\end{equation}
This is shown in Fig. \ref{fig:fig2}(a), where the GP prediction (solid purple line) is perfectly reproduced by that of Eq. (\ref{eq:linearzeta}) (dotted cyan line). In general, if $u_{0}$ is not too large, namely the interaction term does not exceed significantly the kinetic one, also the frequency obtained from the TM model \cite{smerzi1997,leggett2001,pitaevskii2016,burchianti2017}
\begin{equation}
\omega_{J}^{TM}=(2K/\hbar)\sqrt{1+\Lambda}
\end{equation}
can provide a reasonable estimate. In the present case ($\mu/V_{0}=0.25$, $u_{0}=4$), the prediction of the TM model -- that here coincides with that of the fTM model -- exceeds the exact frequency by approximately a $8\%$ ($\omega_{J}=0.595$, $\omega_{J}^{TM}=0.643$), see the dotted green line in Fig. \ref{fig:fig2}(a). This difference may increase further by increasing $u_{0}$ \cite{burchianti2017}. 

\begin{figure}[]
\includegraphics[width=\columnwidth]{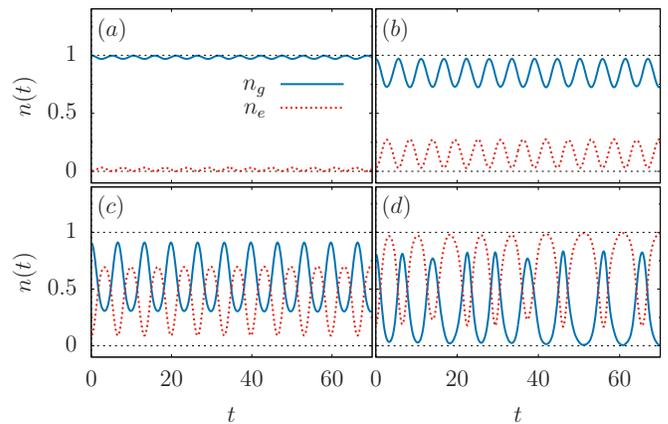}
\caption{Evolution of the occupation number of the ground state and of the Bogoliubov modes, $n_{g}$ and $n_{e}$ respectively, for $z_{0}=0.1$ (a), $0.3$ (b), $0.5$ (c), $0.7$ (d). Notice that in (a), where only the lowest excited mode is appreciably populated, $n_{e}\equiv n_{e_{1}}$, the oscillation period of the population is half of that of the imbalance, namely $T=\pi/\omega_{1}$, see Eq. (\ref{eq:population}).}
\label{fig:occupation}
\end{figure}

In this regime we also have
\begin{equation}
\int\!\! dx |\psi(x,t)|^{2}\simeq n_{g}(t)+n_{e_{1}}(t)
\end{equation},
where $n_{g}(t)\equiv |1+b_{g}(t)|^{2}$ represents the (relative) population of the ground state, and
\begin{align}
n_{e_{1}}(t)&\equiv |b_{10}|^{2} \int\!\! dx \left(|\tilde{u}_{1}|^{2} + |\tilde{v}_{1}|^{2}\right) 
\nonumber\\
&\qquad
+ 2b_{10}^{2}\cos(2\omega_{1}t) \int\!\! dx \tilde{u}_{1}\tilde{v}_{1}
\label{eq:population}
\end{align}
that of the first Bogoliubov excitation, the other excited modes being essentially irrelevant.
The evolution of $n_{g}$ and $n_{e_{1}}$ is plotted in Fig. \ref{fig:occupation}(a) for $z_{0}=0.1$ (the other three panels will be discussed later on).  A sinusoidal oscillation -- with frequency $2\omega_{1}$, see Eq. (\ref{eq:population}) -- is clearly visible in Fig. \ref{fig:occupation}(a). It corresponds to a (small) periodic transfer of population between the ground state and the first excited state, contrarily to what happens in a truly linear system, where the occupation number of each energy level is constant. 

\subsection{Intermediate regime}
\label{sec:interm}

\begin{figure}[t!]
\includegraphics[width=0.8\columnwidth]{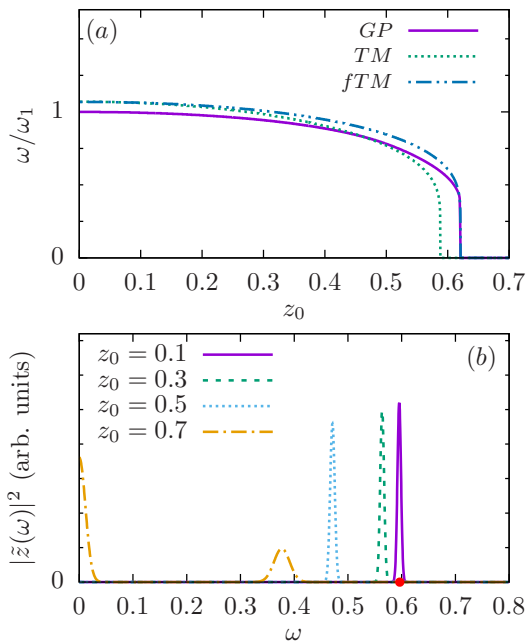}
\caption{(a) Frequency $\omega=2\pi/T$ of the oscillations of $z(t)$ as a function of the initial imbalance $z_{0}$ obtained from the solution of the GP equation and from the TM and the fTM models. (b) Fourier spectrum of $z(t)$ (calculated over an interval of size $t=10^{3}$), for different values of the initial imbalance $z_{0}$. The (red) point on the horizontal axis indicate the value of the first Bogoliubov frequency, $\omega_{1}\simeq 0.6$.  Higher modes lye far outside the present range (e.g. $\omega_{2}\simeq 1.84$, $\omega_{3}\simeq 2.19$).}
\label{fig:fig4}
\end{figure}

In general, when one increases the initial imbalance $z_{0}$, the form of $z(t)$ changes, and its frequency (the inverse of the period) changes as well \cite{smerzi1997}, see Fig. \ref{fig:fig2}(b) and \ref{fig:fig2}(c). In particular, before entering the ST regime, the imbalance is still characterized by periodic oscillations, but with a frequency $\omega$ that is shifted with respect to the plasma value $\omega_{J}$ as an effect of the nonlinearity, see Fig. \ref{fig:fig4}(a). These changes are reflected in the change of the Fourier spectrum, in Fig. \ref{fig:fig4}(b).
Remarkably, in this regime the spectrum is still peaked around a single frequency, that is shifted continuously towards lower frequency values with respect to $\omega_{J}\equiv\omega_{1}$, contrarily to the naive expectation of having more Bogoliubov modes macroscopically occupied (the first Bogoliubov frequencies is indicated by the solid (red) point on the horizontal axis of Fig. \ref{fig:fig4}(b), higher modes lye far outside the present range). In fact, we find that the system exits the linear regime, namely Eq. (\ref{eq:linearcase}) fails in reproducing the actual behavior of $z(t)$ -- see Fig. \ref{fig:fig2}(b) and \ref{fig:fig2}(c) -- even if  higher Bogoliubov modes have an initial population that is still below $1\%$ that of the lowest mode. In other words, the linear approach fails not because some of the other excited modes are \textit{initially} macroscopically occupied (as it would be the case for a truly linear system), but because of the nonlinear mixing during the evolution of the system.

\begin{figure*}[t]
\includegraphics[height=0.7\columnwidth]{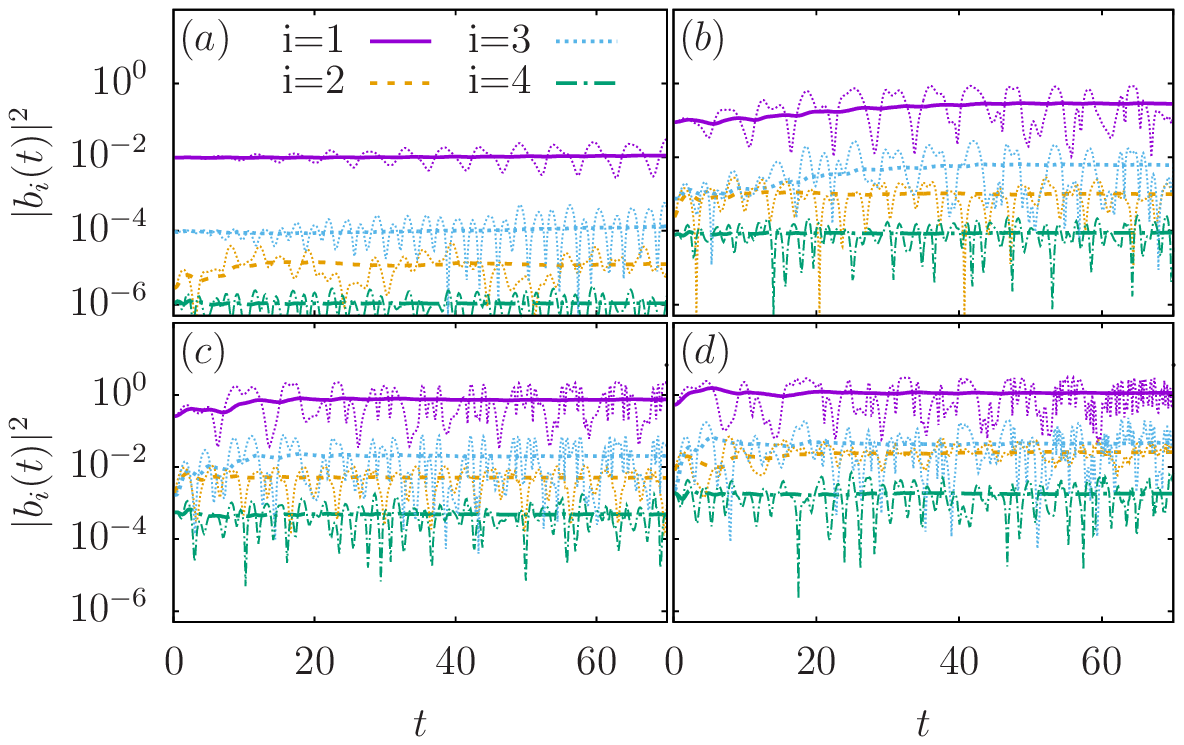}
\hspace{1cm}
\includegraphics[height=0.7\columnwidth]{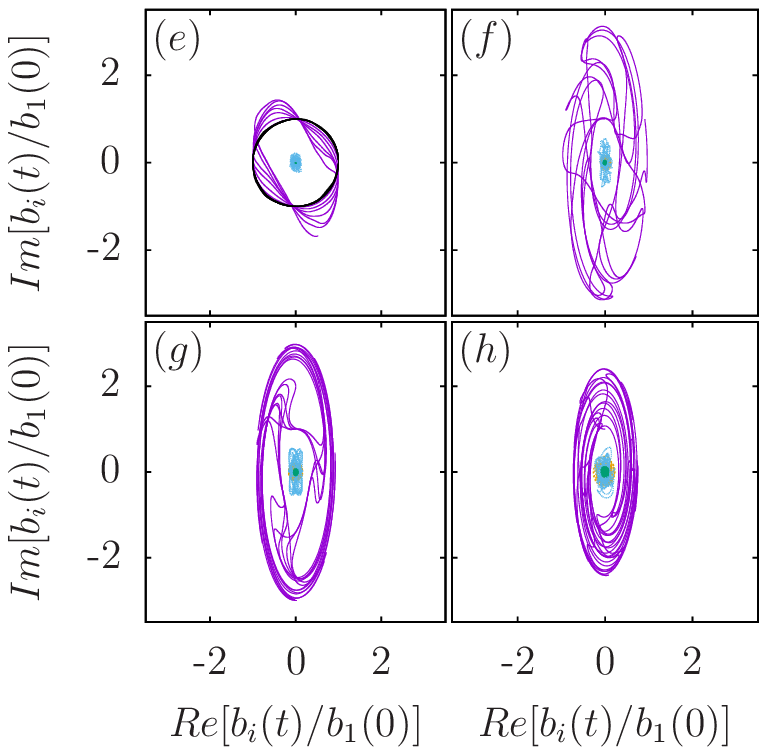}
\caption{(left) Evolution of (the modulus square of) the quasiparticle amplitudes $|b_{i}(t)|^{2}$ ($i=1,2,3,4$, dotted lines, from top to bottom) and of their corresponding time-averaged values $\langle |b_{i}|^{2}\rangle(t)$ [solid lines, see Eq. (\ref{eq:timeaverage})], for $z_{0}=0.1$ (a), $0.3$ (b), $0.5$ (c), $0.7$ (d). (right) Evolution of the real ad imaginary part of $b_{i}(t)$ [normalized to the initial value of $b_{1}(t)$] in the complex plane, for $z_{0}=0.1$ (e), $0.3$ (f), $0.5$ (g), $0.7$ (h). The legends of the two figures are the same. The (black) circle in (e) represents the trajectory for mode 1 in the limit $z_{0}\to0$ (here $z_{0}=0.005$).}
\label{fig:qamplitudes}
\end{figure*}

In this regime, the analog of the decomposition (\ref{eq:population}) becomes more complicated as the contribution of all the excited modes to the total density is now
\begin{align}
\label{eq:ne}
&n_{e}(t)=\sum_{i,j}\left[
b_{i}(t)b_{j}^{*}(t) \int\!\! dx(\tilde{u}_{i}\tilde{u}_{j}^{*} + \tilde{v}^{*}_{j}\tilde{v}_{i}) 
\right.
\nonumber
\\
&
\qquad
\left.
+ b_{i}(t)b_{j}(t) \int\!\! dx\tilde{u}_{i}\tilde{v}_{j} 
+ b_{i}^{*}(t)b_{j}^{*}(t) \int\!\! dx\tilde{u}_{j}^{*}\tilde{v}_{i}^{*} 
\right],
\end{align}
meaning that it is not possible to write the total density as the sum of separate contributions of each Bogoliubov mode. The evolution of $n_{g}(t)$ and $n_{e}(t)$ is shown in Fig. \ref{fig:occupation} for increasing values of the initial imbalance $z_{0}$. This figure shows that the transfer of population between the ground and the excited states increases by increasing $z_{0}$. Initially, when the system exits the linear regime but $z_{0}$ is not too large [e.g. $z_{0}=0.3$, Fig. \ref{fig:occupation}(b)], $n_{g}(t)$ and $n_{e}(t)$ are still characterized by sinusoidal oscillations. For larger values of $z_{0}$, the oscillations in the population deviates from this behavior, as it does the corresponding imbalance [see e.g. Fig. \ref{fig:occupation}(c) and Fig. \ref{fig:fig2}(c)]. In any case, the oscillations of $n_{g}(t)$ are always in phase with those of $|z(t)|$ (that is, the maximal imbalance is obtained when the population of the ground state is maximal). 

\subsection{Self-trapping regime}
\label{sec:st}

By further increasing the initial imbalance $z_{0}$, the period of $z(t)$ gets larger and larger (see also \cite{smerzi1997}), and eventually diverges at the critical value $z_{0c}$ where $\omega\propto1/T\to0$, see Fig. \ref{fig:fig4}(a) ($z_{0c}\simeq0.62$ in the present case). Notice that the value of $z_{0c}$ obtained from the solution of the GP equation is reproduced with great accuracy by the fTM model (we have verified that this holds true even for values of $u_{0}$ larger than that considered in the present paper). Remarkably, the \textit{onset of ST} corresponds to a situation in which the population of the ground state can be transferred completely to the  excited states, namely when $n_{g}(t)=0$ at some $t$ during the evolution, see Fig. \ref{fig:occupation}(d). This feature is indeed a distinctive characteristic of the ST regime. In this regime the imbalance is stuck on the positive side (or the negative one, depending on the initial conditions), still oscillating, but with an irregular pattern \cite{smerzi1997}. The latter reflects in the shape of the frequency spectrum, that significantly broadens and acquires a relevant contribution from the low frequency region, $\omega\simeq 0$ [dotted-dashed orange line in Fig. \ref{fig:fig4}(b)].

\subsection{Non-linear mixing}
\label{sec:nl}
Owing to the coupling between the different Bogoliubov modes, see Eq. (\ref{eq:ne}), we argue that $|b_{i}(t)|^{2}$ cannot be identified with the occupation number of the $i$-th quasiparticle level, contrarily to the the interpretation given in Ref. \cite{morgan1998}. However, since the coefficients $b_{i}(t)$ represent the quasiparticle amplitudes in the sense of the expansion (\ref{eq:exp2}), in the following we shall consider their modulus squared $|b_{i}(t)|^{2}$ as a measure of the weight of each mode in the system dynamics. Their evolution (for $i=1,2,3,4$) is shown in Fig. \ref{fig:qamplitudes}(left), along with the corresponding time-averaged values
\begin{equation}
\langle |b_{i}|^{2}\rangle(t)\equiv\frac{1}{t}\int_{0}^{t}|b_{i}(t')|^{2}dt',
\label{eq:timeaverage}
\end{equation}
for the same values of the initial imbalance as in the previous figures, namely $z_{0}=0.1,0.3,0.5,0.7$. In Fig. \ref{fig:qamplitudes}(right) we show the corresponding region of the complex plane spanned by the real and imaginary parts of $b_{i}(t)$ (here normalized to $b_{1}(0)$, for easiness of visualization) during the evolution of the system. In Fig. \ref{fig:qamplitudes}(e) we also show the trajectory of the lowest Bogoliubov mode ($i=1$) for $z_{0}=0.005$, indicating that in the limit $z_{0}\to 0$ the expected behavior is recovered: in this case the coefficient $b_{1}(t)$ is constant in modulus as dictated by Eq. (\ref{eq:bilinear}) for the linear regime, and the contribution of all higher excited modes is negligible \cite{burchianti2017}. As $z_{0}$ is increased, the dynamics in the complex plane becomes chaotic, each mode spanning a larger portion of the plane. Notice also the change in the orbit shape, from circular (in the limit $z_{0}\to 0$) to elliptical (for $z_{0}\gtrsim0.1$). Both left and right panels evidence a mixing between different modes, especially those with $i=2$ and $i=3$. Remarkably, the mode $i=2$ -- which, being even, does not contribute directly to the imbalance, see Eqs. (\ref{eq:bfull}) and (\ref{eq:imbalance}) -- indeed affects it through the mixing with other modes during the evolution of the system.

\section{Conclusions}
\label{sec:concl}

We have analyzed the dynamics of a (quasi) one-dimensional Bose-Einstein condensate in a double-well potential, from the regime of Josephson plasma oscillations to the self-trapping regime, by means of the Bogoliubov quasiparticle projection method \cite{morgan1998}. In the limit of very small initial imbalance, the system performs Josephson plasma oscillations characterized by the frequency of the lowest Bogoliubov mode (the only Bogoliubov mode   being significantly occupied) \cite{burchianti2017}. In this regime, the evolution of the system is characterized by a  periodic transfer of population between the ground state and the first excited state. As the initial imbalance is increased, the system still performs periodic oscillations between the left and right wells, but with a frequency that is continuously shifted towards values lower than the plasma frequency. 
This occurs because of the nonlinear mixing of the Bogoliubov modes during the evolution of the system, and not because some of the excited modes (besides the lowest one) are initially macroscopically occupied, contrarily to what happens in a linear system. The frequency spectrum of the imbalance is therefore still peaked around a single frequency, and the corresponding period diverges when the system enters the self-trapping regime.
This corresponds to a situation in which the population of the ground state can be transferred completely to the excited states at some time during the evolution. This feature is indeed a distinctive characteristic of the ST regime. The present picture is expected to hold also in higher dimensions.

\begin{acknowledgments}
We thank Alessia Burchianti, Chiara Fort, and Gonzalo Muga for useful discussion during the initial stage of this work. We acknowledge support by the Spanish Ministry of Economy, Industry and Competitiveness and the European Regional Development Fund FEDER through Grant No. FIS2015-67161-P (MINECO/FEDER, UE), and the Basque Government through Grant No. IT986-16. 
\end{acknowledgments}


%

\end{document}